\documentclass[twocolumn,showpacs,secnumarabic,amssymb,superscriptaddress,nobibnotes,aps,pre]{revtex4-2}
\usepackage{amsmath}
\usepackage{amssymb}
\usepackage{graphicx}
\usepackage{url,hyperref}
\usepackage[usenames,dvipsnames]{xcolor}
\hypersetup{colorlinks=true, linkcolor=BrickRed, urlcolor=blue!50!black, citecolor=blue!50!black}
\usepackage[capitalize]{cleveref}
\usepackage{cleveref}
\AtBeginDocument{}
\crefrangelabelformat{equation}{(#3#1#4)$-$(#5#2#6)}
\begin{document}
\title{Should a hotter paramagnet transform quicker to a ferromagnet? Monte Carlo simulation results for Ising model}
\author{Nalina Vadakkayil}
\affiliation{Theoretical Sciences Unit and School of Advanced Materials, 
Jawaharlal Nehru Centre for Advanced Scientific Research, Jakkur P.O., 
Bangalore 560064, India}
\author{Subir K. Das}
\email{das@jncasr.ac.in}
\affiliation{Theoretical Sciences Unit and School of Advanced Materials, 
Jawaharlal Nehru Centre for Advanced Scientific Research, Jakkur P.O., 
Bangalore 560064, India}
\date{\today}
\begin{abstract}
For quicker formation of ice, before inserting inside a refrigerator, 
heating up of a body of water can be beneficial. 
We report first observation of a counterpart of this intriguing fact, referred to as 
the Mpemba effect (ME), during ordering in 
ferromagnets. By performing Monte Carlo simulations of a generic model, we have 
obtained results on relaxation 
of systems that are quenched to sub-critical state points from various temperatures 
above the critical point.  For a fixed final temperature, a system with higher 
starting temperature equilibrates faster 
than the one prepared at a lower temperature, implying the presence of ME. 
The observation is extremely counter-intuitive, 
particularly because of the fact that the model has no in-built frustration or 
metastability that typically is thought to provide 
ME. Via the calculations of nonequilibrium properties concerning structure and energy, 
we quantify the role of critical 
fluctuations behind this fundamental as well as technologically relevant 
observation.
\end{abstract}
\keywords{}
\maketitle
When quenched to the same lower temperature $T_l$, should a hotter system equilibrate faster than a colder one? 
An answer in affirmative is counter-intuitive and relates to the Mpemba effect (ME) \cite{mpemba,jeng}. 
ME can have important applications in memory devices \cite{baity_pnas} and elsewhere. 
In spite of such practical importance and the knowledge since the time of Aristotle \cite{jeng,aristotle,descartes,r_bacon,f_bacon,black}, 
explanation of ME remains elusive. Following the work \cite{mpemba} by Mpemba and Osborne, there has been a surge of interest in understanding it 
\cite{jeng,auerbach,xi,jin,greaney,lu_raz,lasanta,torrente,baity_pnas,klich,gal_raz,chaddah,avinash,burridge}, particularly 
during the last decade \cite{xi,jin,greaney,lu_raz,lasanta,torrente,baity_pnas,klich,gal_raz,chaddah,avinash,burridge}. 
Nevertheless, the progress remains limited. 
Interestingly, there still exists hot debate on the very existence of the effect. Experimental reports are available in favor 
of \cite{auerbach,xi,chaddah,avinash,kuhn} as well as against \cite{burridge,kuhn} it. 

Historically the effect has been attached with cooling or solidification of liquids \cite{aristotle,descartes,r_bacon,f_bacon,zalden,phuah}, like water and milk.
Recently there exist efforts to extend the domain by asking the 
same question for other systems \cite{greaney,lu_raz,lasanta,torrente,baity_pnas,klich,gal_raz,chaddah,avinash}. These include 
cooling granular gases \cite{lasanta}, coarsening spin glasses \cite{baity_pnas}, etc. 
In the case of spin glasses \cite{baity_pnas,binder_young,rieger,lubchenko}, ME is observed due to the variation of the 
correlation length ($\xi$) with the shifting of the starting temperature $T_s$. Likewise, in each type of 
systems \cite{greaney,lasanta,lu_raz,torrente,baity_pnas,klich,gal_raz,chaddah,avinash} 
certain anomaly decides on the existence of ME. 
Some of the studies \cite{baity_pnas,lasanta} provide the impression that the effect has connection with aging systems \cite{fisher_huse,nv_sc_skd,puri}. 
However, it is not clear whether the connection is only with the aging 
systems having glass-like slow dynamics \cite{baity_pnas,binder_young,rieger,lubchenko} or simpler aging systems, undergoing standard clustering or 
phase transitions \cite{bray_adv,onuki}, are also good candidates for the exhibition of the effect.

With the variation of $T_s$ it is expected that certain structural quantities will undergo change. 
In the context of critical phenomena \cite{onuki,fisher}, 
$\xi$ exhibits the divergence \cite{fisher}: $\xi \sim |T_s - T_c|^{-\nu}$, 
as $T_s$ approaches $T_c$, the critical temperature. If variation in quantities associated with structure is responsible for the observation of ME, 
choice of a thermodynamic region close to $T_c$ is then ideal \cite{baity_pnas,saikat_epjb,saikat_pre} for preparing the systems before quenching to a $T_l$. 
Furthermore, to establish the reasons behind the effect, in addition to studies of systems having glass-like 
ingredients, materials of other varieties should also be considered. It is important to study simpler prototype systems. If the effect is observed, such systems 
can provide easier path to understanding, thereby putting the criticisms on the existence of ME to rest.

In this work we consider the standard nearest neighbor {\it{ferromagnetic}} Ising model \cite{fisher,bray_adv,onuki}. 
We explore a wide range of $T_s$, lying above $T_c$. It is convincingly shown that following quenches to a $T_l$, below $T_c$, 
decay of energy for systems with higher $T_s$ occurs faster. This, indeed, is the expectation \cite{baity_pnas} when ME is present. 
Our finding, which we confirmed via various means, is striking, given that, unlike the 
spin glasses, there is no in-built frustrated interaction in this model. 
Observation of ME in such a simple system hints that the 
effect is rather common. Furthermore, we provide a quantitative critical scaling picture to elucidate 
the outcome of the study. 

In the model Hamiltonian \cite{fisher,bray_adv,onuki} $H = -J\sum_{\langle ij \rangle} S_i S_j$, $S_i = \pm 1$ and $J\,(>0)$ 
is the interaction strength between nearest neighbors. The spin values $\pm1$ correspond to up and down orientations of the atomic magnets. 
We study this model via Monte Carlo (MC) simulations \cite{binder_heer,landau,frenkel}, in space dimension $d=2$, on a square lattice. The value of $T_c$ 
for this system \cite{binder_heer} is $\simeq 2.27 J/k_B$, where $k_B$ is the Boltzmann constant.

Following quenches to a $T_l$, the MC simulations were performed by employing the Glauber dynamics \cite{binder_heer,glauber}. 
In this method a trial move is performed by flipping a randomly chosen spin. This does not conserve \cite{bray_adv} the system-integrated order parameter and the dynamics 
corresponds to ordering in a uniaxial ferromagnet. $L^2$ such moves, $L$ being the 
linear dimension of a square simulation box, in units of the lattice constant $a$, make a single MC step (MCS). This is the unit of our time (t).

In the vicinity of $T_c$, the divergence in the relaxation time makes the preparation of initial configurations time taking. To avoid this, we used 
the Wolff algorithm \cite{wolff}, where, instead of a single spin flip, a randomly selected cluster is flipped. Initial configurations prepared at different 
values of $T_s$, via this method, are quenched to several values of $T_l$.

We have applied periodic boundary conditions in both directions. Unless otherwise mentioned, presented results are averaged over 100000 independent initial configurations. All results 
are for $L = 256 a$. In the following we set, for the sake of convenience, $J$, $k_B$ and $a$ to unity.

\begin{figure}
\centering
\includegraphics*[width=0.45\textwidth]{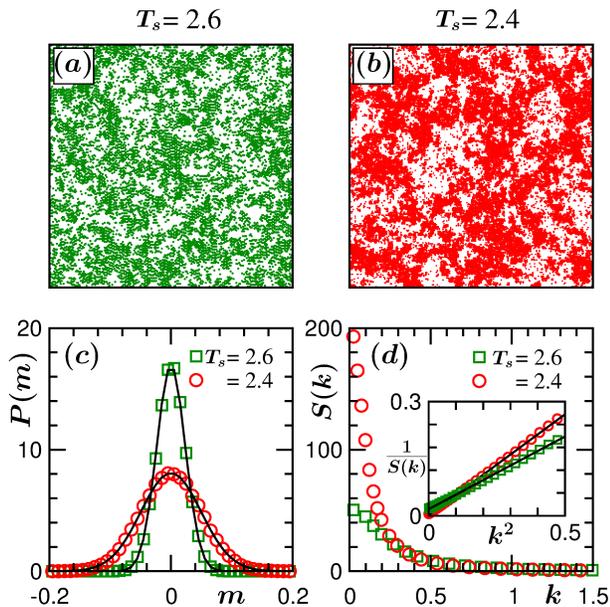}
\caption{\label{fig1}(a)-(b) Typical equilibrium configurations are shown from two starting temperatures, $T_s$. Each of the configurations has 50:50 proportion 
of up and down spins. The locations of the up spins are marked. (c) Plots of equilibrium probability distributions for 
magnetisation are shown from the $T_s$ values for which the snapshots are presented. These results were obtained by exploiting the composition fluctuations in the 
simulations via the Wolff algorithm. The continuous lines are Gaussian fits. 
(d) $S(k)$, the structure factor, is plotted versus the wave number $k$, for 50:50 equilibrium configurations. These results are also presented from the same 
two $T_s$ values as mentioned above. Inset: Plots of $1/S(k)$ versus $k^2$. Here the continuous lines represent linear behavior.}
\end{figure}
In Figs. \ref{fig1}(a) and (b) we show typical equilibrium snapshots from two values of $T_s$, each having critical, i.e., 50:50 proportion of 
up and down spins. The difference in the extent of spatial correlations between the two 
temperatures is easily identifiable from these pictures. Such critical enhancements \cite{fisher} are demonstrated quantitatively in parts (c) and (d) of this figure. 
In Fig. \ref{fig1}(c) we show the probability distributions for order-parameter ($m$) fluctuation \cite{landau,skd_horbach}. The width is much higher for the 
temperature that is closer to $T_c$, implying enhanced susceptibility \cite{fisher,landau,skd_horbach}. In Fig. \ref{fig1}(d) we have presented 
the structure function \cite{fisher}, $S(k)$, versus $k$, the latter being the wave number, for the same two temperatures. 
This structure function is related to the spatial fluctuation in the concentration field when the overall composition is fixed at the critical value \cite{hansen,skd_horbach,skd_horbach2}. 
In the $k \rightarrow 0$ limit, stronger enhancement in $S(k)$, for $T_s$ closer to $T_c$, is again related to higher susceptibility. In the inset of this 
frame we show $1/S(k)$ as a function of $k^2$, in a small $k$ regime. Linear appearances are consistent with the Ornstein-Zernike \cite{fisher,hansen} behavior. 
Steeper slope for smaller $T_s$ signifies an enhancement \cite{fisher,skd_horbach2} in $\xi$ with the approach to $T_c$. With such temperature dependent initial configurations, 
we study the {\it{equilibration}} dynamics following quenches to various $T_l$ below $T_c$.

\begin{figure}
\centering
\includegraphics*[width=0.5\textwidth]{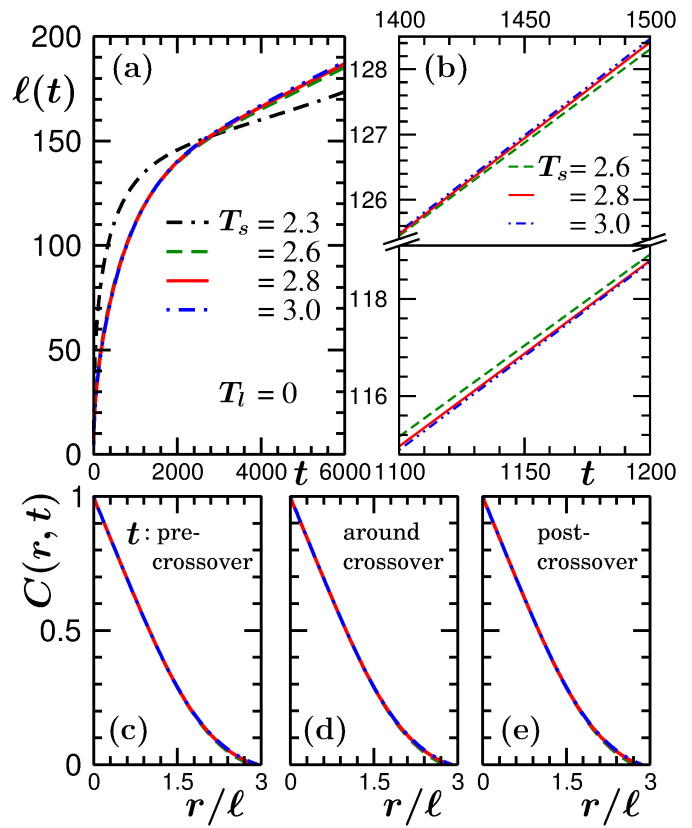}
\caption{\label{fig2} (a) Average domain lengths, following quenches to $T_l = 0$, 
from different values of $T_s$, are plotted 
versus time. (b) Same as (a). Here we have enlarged plots for three of the 
considered $T_s$ values. The broken frame is adopted to bring clarity on the differences among 
early as well as late time data sets. (c)-(e) Two-point equal time correlation functions, 
from different $T_s$, as used in (b), for three different times, are plotted versus $r/\ell$, 
for quenches to $T_l = 0$. See text for details.}
\end{figure}
\begin{figure}
\centering
\includegraphics*[width=0.45\textwidth]{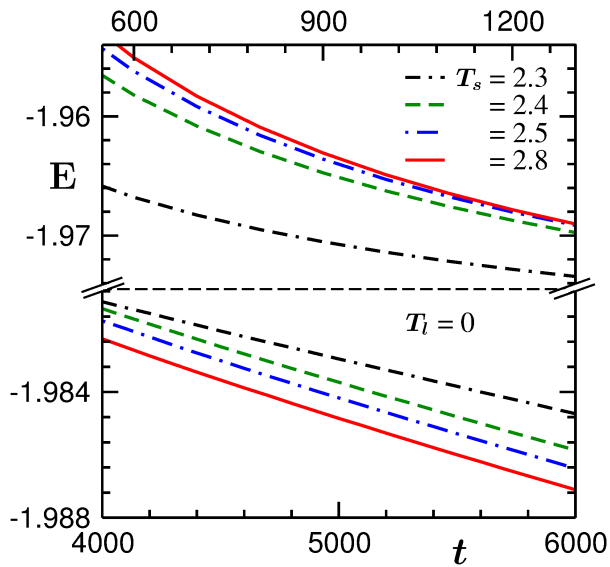}
\caption{\label{fig3} Energy per spin, $E$, is plotted as a function of time. Results from several different choices of $T_s$ are shown. 
In each of these cases the systems were quenched to $T_l = 0$. The frame has been broken to bring clarity in both early time and late time 
trends in the data sets.}
\end{figure}
In Fig. \ref{fig2}(a) we present plots for the growth of average domain length ($\ell$), following quenches of initial configurations prepared 
at different values of $T_s$. In each of these cases the value of $T_l$ was set at zero. These lengths were calculated from the first moment of the 
domain size distribution function \cite{majumder_skd}, size of {\it{a domain}} being estimated 
as the distance between two successive interfaces, while scanning along 
different Cartesian directions. 
It is clearly seen in Fig. \ref{fig2}(a) that there exist crossings among curves and for 
a lower value of $T_s$ the late time average domain lengths are smaller than those for a 
higher $T_s$. This suggests that the systems starting from higher $T_s$ are relaxing faster.

A requirement for the validity of the above discussed picture, on the growths of lengths, 
is the existence of the self-similar property among 
the evolving domains \cite{bray_adv}, for different $T_s$, at any given instant of time 
within 
the relevant period. This feature should get reflected 
in the {\it{simple}} scaling property \cite{bray_adv}, $C(r,T_s) \equiv \tilde{C} (r/\ell)$, 
of the two-point equal time correlation function, 
$C(r,T_s) = \langle S_i S_j \rangle - \langle S_i \rangle \langle S_j \rangle$. Here $r$ is 
the scalar distance between the points $i$ and $j$, while 
$\tilde{C} (x)$ is a master function that should be independent of $T_s$. 
We intend to demonstrate the validity of the above mentioned 
scaling property for times below, equal to, and greater than the crossing times. 
For this purpose in Fig. \ref{fig2}(b) we have shown enlarged plots for a subset of 
$T_s$ values considered in Fig. \ref{fig2}(a) [broken frames are used to bring 
clarity in both early and late time data sets]. From this figure it appears that the 
crossings among these length data sets occur around $t = 1300$. Thus, we have shown the 
scaling plots for the correlation functions, in 
parts (c), (d) and (e) of Fig. \ref{fig2}, for $t = 1100$, $1300$ and $1500$. 
In each of the cases good collapse of data can be observed. 
This fact states that comparisons among length data from different $T_s$ are meaningful. 
Here it is worth mentioning that the initial configurations 
with large enough spatial 
correlation is fractal in nature \cite{fisher}. In that case 
the scaling at early enough times should follow the 
form \cite{fisher,vicsek,midya_skd,paul_skd} 
$C(r,T_s) \equiv r^{d-d_f}\tilde{C}(r/\ell)$, where $d_f$ is the fractal dimension. 
This is consistent with the Ornstein-Zernike form \cite{fisher,hansen} $r^{-p}e^{-r/\xi}$. 
However, the observation of good data collapse for $d_f = d$ implies that the 
fractal features practically disappeared well before the crossing times. 
In the following we present results on energy decay. This is an 
alternative route for the confirmation of ME \cite{baity_pnas}.

\begin{figure}
    \centering
    \includegraphics*[width=0.45\textwidth]{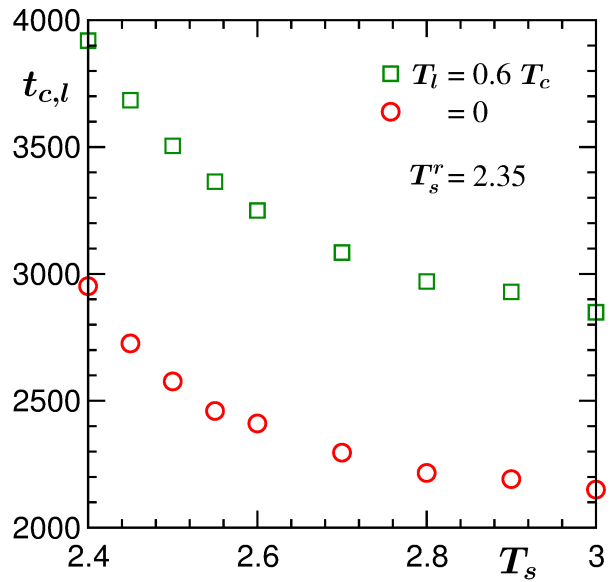}
    \caption{Plots of crossing times, $t_{c,l}$, versus $T_s$, of energy curves for different $T_s$ values with that of a lower reference  starting temperature, viz., 
    $T_s^r = 2.35$. We have shown data for two values of $T_l$, viz., $T_l =0$ and $T_l = 0.6 T_c$.}
    \label{fig4}
\end{figure}

In Fig. \ref{fig3} we show the time dependence of $E$, energy per spin, during evolutions for different $T_s$ values, again by fixing $T_l$ to zero. Like in the first frame of Fig. \ref{fig2}, 
there exist crossings here as well. Systems with higher $T_s$, i.e., larger starting energy, are approaching new equilibrium faster. This, indeed, is the 
basic essence of ME. The crossings are very systematic. This is owing to extremely good statistics. A better quantitative information on the trend of crossings is demonstrated in Fig. \ref{fig4}. 
Here we have plotted the crossing times, $t_{c,l}$, of energy curves for different values of $T_s$, following quenches to a $T_l$, 
with that for a reference value $T_s^r = 2.35$. We have shown results for $T_l = 0$ and $0.6 T_c$. 
Each of these data sets conveys the message that energy plots for higher 
values of $T_s$ are crossing the reference plot earlier. This indirectly implies that there exists crossing between any chosen pair of curves. 
This required feature is present in the plots for both the values of $T_l$. 

\begin{figure}
    \centering
    \includegraphics*[width=0.45\textwidth]{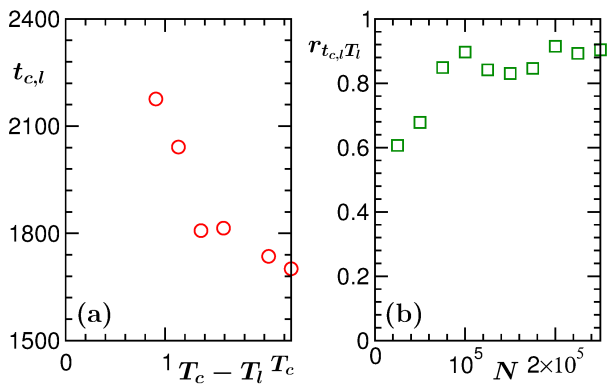}
    \caption{(a) Plot of $t_{c,l}$, as a function of $T_c - T_l$, 
    for crossings between energy curves from two $T_s$ values, viz., $2.5$ and $2.6$. The data set is presented after averaging over 200000 independent 
    initial realizations.
    (b) Pearson correlation coefficient, $r_{t_{c,l}T_l}$, is shown with the variation of $N$, the number of initial configurations used in the averaging.}
    \label{fig5}
\end{figure}

A comparison between the two plots in Fig. \ref{fig4} suggests that with the increase 
of $T_l$ crossing between curves for two different $T_s$ values has become delayed. 
This may imply that the crossing time will diverge with the approach of $T_l$ towards $T_c$. A comprehensive exercise related to that is shown in 
Fig. \ref{fig5}(a). 
Here $t_{c,l}$ represent the crossing times between the energy curves for $T_s = 2.5$ and $2.6$, following quenches to different $T_l$. 
The trend is consistent with the above anticipated singularity and points to a possibility that 
a phase transition is necessary to observe the ME, i.e., $T_s$ and $T_l$ should lie on two sides of the critical point. However, for a concrete statement on the latter 
independent studies are needed by fixing $T_s$ and $T_l$ on the same sides of $T_c$.

To further ascertain the effects of critical fluctuations at $T_l$, on the magnitude of ME, we present additional results in Fig. \ref{fig5}(b). 
Given the debates on the topic, while good statistics is a necessity, it is also essential to demonstrate that there exists no bias in the presented results due to the averaging over a 
specific set of initial configurations. Keeping that in mind we have calculated the Pearson correlation 
coefficient \cite{pearson}, 
$r_{t_{c,l}T_l} = \sum_{l=1}^n x_l y_l /\sqrt{\sum_{l=1}^n x_l^2} \sqrt{\sum_{l=1}^n y_l^2}$. 
Here $x_l$ and $y_l$ are, respectively, $t_{c,l}-\overline{t}_{c}$ and $T_l - \overline{T}$, with $\overline{t}_{c}$ and $\overline{T}$ being averages of crossing times and quench temperatures 
for a sample of size $n$ ($=6$ here). We have calculated this coefficient by using $t_{c,l}$ versus $T_l$ data that were obtained by averaging over increasing number ($N$) of initial configurations. 
The $r_{t_{c,l}T_l}$ versus $N$ plot in Fig. \ref{fig5} (b) clearly conveys the message that the correlation between $t_{c,l}$ and $T_l$ is positive, thereby discarding the possibility of 
aforementioned biasness unambiguously.

We have studied kinetics of phase transitions \cite{bray_adv} in the two-dimensional nearest neighbour Ising model \cite{fisher}, via Monte Carlo simulations \cite{binder_heer}, 
using Glauber spin-flip dynamics \cite{glauber}. This mimics the ordering dynamics in uniaxial ferromagnets. 
The objective has been to investigate the presence of Mpemba effect \cite{mpemba}. 
For this purpose we have prepared 
initial configurations at various starting temperatures $T_s$, lying between $T_c$ and $\infty$. These configurations, with 50:50 compositions 
of up and down spins, were quenched to various final temperatures $T_l$ ($<T_c$). We observe that systems with higher $T_s$ tend to approach the 
equilibrium at a $T_l$ faster than the ones with lower $T_s$. This is the basic fact of Mpemba effect \cite{mpemba}.

While Mpemba effect itself is a counter-intuitive phenomena, observation of it in a simple system that is considered here is even more 
surprizing. Note that the model has no glassy ingredient. 
We have presented results for multiple values of $T_l$. In each of the cases, the effect is clearly 
identifiable. 
We have also shown that as $T_l$ increases towards $T_c$, the 
crossing time between energy curves for a pair of starting temperatures increases. 
This may imply that a phase transition is necessary for the observation of 
ME. However, further studies are necessary to arrive at such a conclusion.

Despite no in-built glassy feature, the model has been recognized 
\cite{nv_sc_skd,redner,cugliandolo} to exhibit unusual structure and slow dynamics 
at $T_l = 0$, particularly in $d=3$. Such behavior may be considered to be a reason behind our striking observation. Nevertheless, interestingly, the effect 
is also observed for much higher values of $T_l$ and in $d=2$. Our results suggest that it persists at least till $T_l$ is less than $T_c$. This work, thus, we expect to inspire further 
novel investigations, experimental as well as theoretical, with simple systems, providing path towards better understanding of the Mpemba effect.

In this work we have considered initial configurations with 50:50 
compositions of up and down spins. It is equally important to study the case of 
asymmetric starting compositions. In this case also variation of the correlation length 
in the starting configurations can be realized with the change in temperature. Thus, the 
effect may be observed for non-zero initial magnetization as well. Our preliminary 
studies support this expectation. Nevertheless, more thorough 
studies are needed. Here we have considered the Glauber dynamics \cite{glauber} for 
which the order parameter does not remain conserved over time. 
It will be interesting to extend the investigation to the 
conserved order-parameter dynamics via the implementation of Kawasaki exchange 
kinetics \cite{landau}. A systematic study of this, however, can be time taking. 
Note that in the case of Kawasaki kinetics, due to significantly slower 
growth \cite{majumder_skd} the crossings may occur at much later times.
\section*{Author contributions}
SKD conceived the project, designed the problem, supervised the work and wrote 
the manuscript. NV wrote the computer codes, carried out the work and contributed 
to the preparation of the manuscript.
\section*{Conflicts of interest}
There are no conflicts to declare.
\section*{Acknowledgements}
SKD acknowledges encouraging remarks from or discussions with K. Binder, C.N.R. Rao, M. Zannetti, R. Pandit, C. Dasgupta, P. Chaudhuri, S. Sastry, J.K. Bhattacharjee, K. Sengupta and 
J. Horbach. The authors received partial 
financial support from Science and Engineering Research 
Board, India, via Grant No. MTR/2019/001585. They are thankful to a Supercomputing 
facility at JNCASR.

\end{document}